\documentclass[12pt]{article}
\usepackage{amsmath,amssymb,epsfig,subfigure}

\renewcommand{\le}{\leqslant}
\renewcommand{\ge}{\geqslant}
\newcommand{\be}{\begin{equation}}
\newcommand{\en}{\end{equation}}

\newcommand{\ii}{\textrm{i}}
\newcommand{\ee}{\textrm{e}}
\renewcommand{\vec}[1]{\boldsymbol{#1}}

\newcommand{\la}{\label}

\def\rr#1{(\ref{#1})}

\def \div{\mbox{div\hskip 1pt}}
\def \Div{\mbox{Div\hskip 1pt}}
\def \tr{\mbox{tr\hskip 1pt}}
\def \grad{\mbox{grad\hskip 1pt}}
\def \Grad{\mbox{Grad\hskip 1pt}}

\def\bm#1{\mbox{\boldmath{$#1$}}}

\begin{document}

\numberwithin{equation}{section}

%++++++++++++++++++++++++++++++++++++++++++++++++++=++++++++++++++++++

\title{Acoustic waves at the interface of \\a pre-stressed incompressible
elastic solid \\and a viscous fluid}
%++++++++++++++++++++++++++++++++++++++++++++++++++++++++++++++++++++
\author{M. Ott\'enio, M. Destrade, R.W. Ogden}
\date{2006}
\maketitle
%
%++++++++++++++++++++++++++++++++++++++++++++++++++++++++++

%++++++++++++++++++++++++++++++++++++++++++++++++++++++++++
\begin{abstract}

We analyze the influence of pre-stress on the propagation of
interfacial waves along the boundary of an incompressible
hyperelastic half-space that is in contact with a viscous fluid
extending to infinity in the adjoining half-space.

One aim is to derive rigorously the incremental boundary
conditions at the interface; this derivation is delicate because
of the interplay between the Lagrangian and the Eulerian
descriptions but is crucial for numerous problems concerned with
the interaction between a compliant wall and a viscous fluid. A
second aim of this work is to model the ultrasonic waves used in
the assessment of aortic aneurysms, and here we find that for this
purpose the half-space idealization is justified at high
frequencies. A third goal is to shed some light on the stability
behaviour in compression of the solid half-space, as compared with
the situation in the absence of fluid; we find that the usual
technique of seeking standing waves solutions is not appropriate
when the half-space is in contact with a fluid; in fact, a correct
analysis reveals that the presence of a viscous fluid makes a
compressed neo-Hookean half-space slightly more stable.

For a wave travelling in a direction of principal strain, we
obtain results for the case of a general (incompressible
isotropic) strain-energy function. For a wave travelling parallel
to the interface and in an arbitrary direction in a plane of
principal strain, we specialize the analysis to the neo-Hookean
strain-energy function.

\end{abstract}

%

%+++++++++++++++++++++
\section{Introduction}
%+++++++++++++++++++++

Seismic records show that underground rocks and ocean beds are
subject to stress and strain and that surrounding fluids are
viscous and under high pressures. Clinical ultrasonic measurements
indicate that arteries can undergo large strains in service and
are most sensitive to changes in blood pressure. Many moving and
vibrating parts of automotive devices are made of loaded
elastomers in contact with highly viscous fluids. These are a few
examples of situations where it is crucial to model and understand
the motions and the stability of the interface between a
\emph{deformed} elastic solid and a \emph{viscous} fluid. Yet only
a handful of studies can be found on the subject, especially when
compared with the abundant literature on waves at the interface
between an elastic solid and an \emph{inviscid} fluid, which goes
from the pioneering works of Galbrun \cite{Galbrun}, Cagniard \cite{Cagniard},
Scholte \cite{Scholte}, and Biot \cite{Biot52} to the definitive treatment of the
acoustoelastic effect by Sinha et al. \cite{Sinha}; see also Poir\'ee
\cite{Poir} and Degtyar and Rohklin \cite{Degtyar}. Waves at the interface
between a viscous fluid and an \emph{undeformed} isotropic elastic
solid were examined by Vol'kenshtein and Levin \cite{Vol} and the
corresponding problem for an anisotropic elastic solid by Wu and
Wu \cite{Wu}. To the best of our knowledge, only Bagno, Guz, and
their co-workers have studied the title problem (see, for example,
\cite{Bagno87}, \cite{Bagno97}). Their analytical treatment is,
however, quite succinct and we therefore aim to shed new light on
the problem by re-examining it on the basis of recent developments
in the theory of small-amplitude waves, linearized in the
neighbourhood of a finite, static, homogeneous deformation.

It turns out that one of the trickiest aspects of the study is the
derivation of proper incremental boundary conditions at the
interface because these are usually written in terms of the
nominal stress in a deformed solid (Lagrangian formulation,
Section 2) and in terms of the Cauchy stress in a fluid (Eulerian
formulation, Section 3). These equations are combined in an
appropriate way for a general interface in Section 4. We then
specialize the analysis to principal wave propagation for an
arbitrary (incompressible, isotropic) strain-energy function in
Section 5. In the course of the analysis in Section 5, by way of
application of the theory, we show that in respect of an abdominal
aortic aneurysm it is appropriate to neglect the curvature and
finite thickness for ultrasonic waves (10\,MHz), i.e. to treat the
aneurysm locally as a half-space. It also shown that it is not
appropriate to use waves with a real frequency to study the
stability of compressed solids in contact with a viscous fluid.
Finally, in Section 6, by specializing to the neo-Hookean solid,
we consider the propagation of non-principal waves for both
tension and compression of the half-space in order to illustrate
the influence of the fluid.

%++++++++++++++++++++++++++++++++++++
\section{Basic equations for the solid}
%++++++++++++++++++++++++++++++++++++

For the solid material we denote by $\vec{F}$ the deformation
gradient relating the stress-free reference configuration, denoted
$\mathcal{B}_0$, to the finitely deforming configuration, denoted
$\mathcal{B}$. This has the form $\vec{F}=\Grad\vec{x}$, where
$\vec{x}=\bm{\chi}(\vec{X},t)$ is the position vector in
$\mathcal{B}$ at time $t$ of a material point located at $\vec{X}$
in $\mathcal{B}_0$, $\bm{\chi}$ is the deformation mapping, and
$\Grad$ is the gradient operator relative to $\mathcal{B}_0$.

We consider the material to be elastic with a strain-energy
function, defined per unit volume, denoted by $W=W(\vec{F})$.
Furthermore, we restrict attention to incompressible materials so
that the constraint
\begin{equation}
\det\vec{F}=1\label{constraint1}
\end{equation}
is in force. The nominal stress tensor, here denoted by $\vec{S}$,
and the Cauchy stress tensor $\bm{\sigma}$ are then given by
\begin{equation}
\vec{S}=\frac{\partial W}{\partial\vec{F}}-p\vec{F}^{-1},\quad
\bm{\sigma}=\vec{F}\frac{\partial
W}{\partial\vec{F}}-p\vec{I},\label{nominal-cauchy}
\end{equation}
where $p$ is a Lagrange multiplier associated with the constraint
\eqref{constraint1} and $\vec{I}$ is the identity tensor.

The equation of motion is
\begin{equation}
\Div\vec{S}=\rho{\bf\ddot{\vec{x}}},\label{motion1}
\end{equation}
where $\Div$ is the divergence operator relative to $\mathcal{B}_0$,
$\rho$ is the mass density of the material, and a superposed dot
signifies the material time derivative.

Next, we consider a small motion superimposed on the finite
deformation. Let $\vec{u}(\vec{x},t)$ be the displacement vector
relative to $\mathcal{B}$ and
$\vec{v}(\vec{x},t)={\bf\dot{\vec{u}}}$ the associated particle
velocity (the material time derivative of $\vec{u}$). Then, on
taking the increment of equation \eqref{motion1} and thereafter
changing the reference configuration from $\mathcal{B}_0$ to
$\mathcal{B}$, we obtain
\begin{equation}
\div\vec{s}=\rho{\bf\ddot{\vec{u}}}\equiv\rho{\bf\dot{\vec{v}}},
  \label{motion2}
\end{equation}
where $\vec{s}$ is the increment in $\vec{S}$ (referred to
$\mathcal{B}$) and div the divergence operator relative to
$\mathcal{B}$.

The (linearized) incremental version of the constitutive relation
\eqref{nominal-cauchy} is written
\begin{equation}
\vec{s} =\bm{\mathcal{A}}(\grad\vec{u})
+p(\grad\vec{u})-\tilde{p}\vec{I},
\end{equation}
where $\bm{\mathcal{A}}$ is a fourth-order tensor of elastic moduli,
$\grad$ is the gradient operator relative to $\mathcal{B}$, and
$\tilde{p}$ is the increment in $p$.  In component form, this is
written
\begin{equation}
s_{ij}=\mathcal{A}_{ijkl}u_{l,k}+pu_{i,j}-\tilde{p}\delta_{ij},
\end{equation}
where $_{,k}$ denotes $\partial/\partial x_k$ and $\delta_{ij}$ is
the Kronecker delta.  In terms of $W$ the components of
$\bm{\mathcal{A}}$ are given by
\begin{equation}
\mathcal{A}_{ijkl}=F_{ip}F_{kq}\frac{\partial^2W}{\partial
F_{jp}\partial F_{lq}}.
\end{equation}
For details of these derivations (in a slightly different
notation) we refer to Dowaikh and Ogden \cite{Dowaikh90}.

We now consider the material to be isotropic, so that
$W=W(\lambda_1,\lambda_2,\lambda_3)$ is a symmetric function of the
principal stretches, $\lambda_1,\lambda_2,\lambda_3$ (the positive
square roots of the principal values of $\vec{F}\vec{F}^T$, where
$^T$ signifies the transpose), subject to the constraint
\begin{equation}
\lambda_1\lambda_2\lambda_3=1,\label{constraint2}
\end{equation}
which follows from \eqref{constraint1}. Then, on noting that for
an isotropic material $\bm{\sigma}$ is coaxial with
$\vec{F}\vec{F}^T$ and specializing equation
\eqref{nominal-cauchy}$_2$, we obtain the principal Cauchy
stresses (see, for example, Ogden \cite{Ogden})
\begin{equation}
\sigma_i  = -p + \lambda_i W_i, \quad i =1, 2, 3\quad \text{(no sum
over $i$)},\label{principal-cauchy}
\end{equation}
where $W_i=\partial W / \partial \lambda_i,\,i=1,2,3$.

When referred to the same principal axes, the only non-zero
components of $\bm{\mathcal{A}}$ are
\begin{eqnarray}
&&\mathcal{A}_{iijj}=\lambda_i\lambda_jW_{ij},\quad
i,j\in\{1,2,3\},\\[0.2cm]
&&\mathcal{A}_{ijij}
 =\frac{\lambda_iW_i-\lambda_jW_j}
            {\lambda_i^2-\lambda_j^2}\lambda_i^2,\quad
i,j\in\{1,2,3\},\, i\neq j,\\[0.2cm]
&&\mathcal{A}_{ijji}=
  \mathcal{A}_{jiij}=\mathcal{A}_{ijij}-\lambda_iW_i,\quad
i,j\in\{1,2,3\},\, i\neq j,
\end{eqnarray}
where $W_{ij}=\partial^2W/\partial\lambda_i\partial\lambda_j$. For
subsequent convenience, we adopt the notations defined by
\begin{align}
& \gamma_{ij} =
  (\lambda_i W_i-\lambda_j W_j)\lambda_i^2/(\lambda_i^2-\lambda_j^2),
  \nonumber \notag
  \\[0.1cm]
& \beta_{ij} = (\lambda_i^2 W_{ii} + \lambda_j^2 W_{jj})/2
    - \lambda_i \lambda_j W_{ij} + (\lambda_i W_j-\lambda_j W_i)
     \lambda_i \lambda_j/(\lambda_i^2-\lambda_j^2),
 \label{gammaBeta}
\end{align}
noting that $
\gamma_{ji} \lambda_i^2 = \gamma_{ij} \lambda_j^2$ and $\beta_{ji}=\beta_{ij}$.

%++++++++++++++++++++++++++++++++++++
\subsection{The pre-stressed elastic half-space}
%++++++++++++++++++++++++++++++++++++

We now consider $\mathcal{B}$ to be independent of time and to
correspond to a pure homogeneous strain of a half-space defined by
$x_2 \ge 0$.  The half-space is maintained in this configuration
so that its boundary $x_2=0$ is a principal plane of strain. We
denote by $x_1$ and $x_3$ the other two principal directions of
strain and by $\lambda_1$, $\lambda_2$, $\lambda_3$ the principal
stretches in the $x_1$, $x_2$, $x_3$ directions, respectively. The
corresponding principal Cauchy stresses are then as given by
\eqref{principal-cauchy}.  In particular, the boundary $x_2=0$ is
subject to a normal stress $\sigma_2$ and, after elimination of
$p$, the other two principal Cauchy stresses are then given by
%\begin{figure}
% \centering
%  \epsfig{figure=fig1.eps, height=.5\textwidth, width=.8\textwidth}
%\caption{Geometry of the layered structure}
%\label{Figure1}
%\end{figure}
\begin{equation} \label{initialStress}
\sigma_1 = \sigma_2 + \lambda_1 W_1 - \lambda_2 W_2, \quad \sigma_3
=  \sigma_2 +  \lambda_3 W_3 - \lambda_2 W_2.
\end{equation}

We are interested in the propagation of incremental (small
amplitude) acoustic waves along the boundary plane $x_2=0$, in a
direction making an angle $\theta$ with the principal direction
$x_1$.  The incremental velocity and nominal stress fields
$\vec{v}$ and $\vec{s}$ are then considered as superimposed on
this finite static configuration.  We examine inhomogeneous
time-harmonic plane waves of the form
\begin{equation}
\{ \vec{v}, \vec{s} \}(x_1, x_2, x_3, t)
  = \{ \hat{\vec{v}}(x_2), -(k/\omega) \hat{\vec{s}}(x_2)\}
                \ee^{\ii k(c_\theta x_1 + s_\theta x_3)}
                  \ee^{- \ii \omega t},\label{wave}
\end{equation}
where  we have introduced the notations $c_\theta = \cos\theta$,
$s_\theta = \sin\theta$, $k$ is the wave number,
$\omega$ is the wave frequency, and $\hat{\vec{v}}$,
$\hat{\vec{s}}$ are functions of $x_2$ only, such that
\begin{equation}
 \hat{\vec{v}}(\infty) = \mathbf{0}, \quad \hat{\vec{s}}(\infty)
   = \mathbf{0}.
\end{equation}
Using the results of Destrade et al. \cite{Destrade} (see also Chadwick
\cite{Chadwick}), we find that the incremental equations of motion can be
written as a first-order differential system of six equations,
namely
\begin{equation}
  \vec{\xi}'(x_2) = \ii k \vec{N}\vec{\xi}(x_2),\label{1stOrder}
\end{equation}
where the notation $\vec{\xi}$ is defined by
\begin{equation}
 \vec{\xi} =
    [\hat{v}_1, \hat{v}_2, \hat{v}_3,
      \hat{s}_{21}, \hat{s}_{22}, \hat{s}_{23}]^T,
\end{equation}
and the $6 \times 6$ matrix $\vec{N}$ has the block structure
\begin{equation}
\vec{N}  =  \begin{bmatrix}
                    \vec{N}_1\ & \vec{N}_2 \\[0.1cm]
   \vec{N}_3 + \hat{\rho} \vec{I}\ \  & \vec{N}_1^T
                             \end{bmatrix},\label{N}
\end{equation}
in which the $3 \times 3$ matrices $\vec{N}_1$, $\vec{N}_2$,
$\vec{N}_3$ are real and their components depend on the material
parameters $\gamma_{ij}$ and $\beta_{ij}$ given in
\eqref{gammaBeta}, and the notation $\hat{\rho}=\rho \omega^2 /
k^2$ has been introduced. Here $\vec{I}$ represents the $3 \times
3$ identity matrix.

Explicitly, $-\vec{N}_1$, $\vec{N}_2$, $- \vec{N}_3$ are
\begin{equation}
 \begin{bmatrix}
       0 & c_\theta (\gamma_{21}-\sigma_2)/\gamma_{21} & 0 \\
       c_\theta & 0 & s_\theta \\
       0 & s_\theta (\gamma_{23}-\sigma_2)/\gamma_{23} & 0
       \end{bmatrix},
\quad
 \begin{bmatrix}
       1/\gamma_{21} & 0 & 0 \\
           0 & 0 & 0 \\
        0 & 0 &  1/\gamma_{23}
      \end{bmatrix},
\quad
 \begin{bmatrix}
                       \eta   &     0     & -\kappa \\
                          0   &     \nu   &      0     \\
                    - \kappa  &     0     &  \mu
                   \end{bmatrix},\label{N1N2N3}
\end{equation}
respectively, where
\begin{eqnarray}
&& \eta =
    2 c_\theta^2(\beta_{12} + \gamma_{21} - \sigma_2)
          + s_\theta^2 \gamma_{31},
\nonumber \\[0.1cm]
&& \nu =
  c_\theta^2[\gamma_{12}
      - (\gamma_{21}-\sigma_2)^2 / \gamma_{21}]
   + s_\theta^2[\gamma_{32}
            - (\gamma_{23}-\sigma_2)^2/\gamma_{23}],
\nonumber \\[0.1cm]
&& \mu =
    c_\theta^2 \gamma_{13}
       +  2 s_\theta^2 (\beta_{23} + \gamma_{23} - \sigma_2),
\nonumber \\[0.1cm]
&& \kappa
    =   c_\theta s_\theta (\beta_{13} - \beta_{12} - \beta _{23}
                             - \gamma_{21} - \gamma_{23} + 2\sigma_2).
\end{eqnarray}
Equation \eqref{1stOrder} provides the general expression for the
equations of motion, for arbitrary $\theta$ and $W$.

Now, in seeking a decaying partial-mode solution of the form
\begin{equation}
 \vec{\xi}(x_2) = \ee^{- k s x_2}\vec{\zeta},
 \quad \Re(k s) >0,\label{solutionSolid}
\end{equation}
where $\vec{\zeta}$ is a constant vector and $s$ an unknown
scalar, we arrive at the eigenvalue problem $\vec{N}\vec{\zeta} =
\ii s \vec{\zeta}$. In general, the associated \textit{propagation
condition\/}, $\det (\vec{N} - \ii s \vec{I}) =0$, is a cubic in
$s^2$  \cite{Rogerson}, where now $\vec{I}$ is the
$6\times 6$ identity matrix. Its analytical resolution is too
cumbersome to be of practical interest, and so we specialize the
general equations to the following, simpler, situations: (i)
principal wave propagation ($\theta = 0$) for arbitrary $W$; (ii)
non-principal wave propagation ($\theta \ne 0$) for the
neo-Hookean material, for which
\begin{equation}
W = C(\lambda_1^2 + \lambda_2^2 +
\lambda_3^2-3)/2,\label{neoHookean}
\end{equation}
where $C >0$ is a constant (the shear modulus of the material in
the reference configuration).

In Case (i), the equations of motion decouple the system
\begin{equation}
\begin{bmatrix}
       \hat{v}_3'\\
       \hat{s}_{23}'
\end{bmatrix}
 = \ii k \begin{bmatrix}
       1/\gamma_{21} & 0 \\
       0 &  1/\gamma_{23}
      \end{bmatrix}
\begin{bmatrix}
       \hat{v}_3\\
       \hat{s}_{23}
\end{bmatrix},
\end{equation}
(for which the trivial solution may be chosen) from a system of four
differential equations for $\hat{v}_1$, $\hat{v}_2$, $\hat{s}_{21}$,
$\hat{s}_{22}$. Hence, in this case, the wave is elliptically
polarized, in the $(x_1,x_2)$ plane. The corresponding propagation
condition is a quadratic in $s^2$, which can be solved explicitly.

In Case (ii), we also find that the wave is two-partial, polarized
in the plane containing the directions of propagation and
attenuation (the saggital plane); there, the corresponding
propagation condition involves the product of the factor ($s^2 -
1$) and a term linear in $s^2$, which simplifies the analysis.

Before embarking on the details of these cases, we complete the
description of the boundary-value problem by considering the
behaviour of the wave in the fluid in the half-space $x_2 \leq 0$.

%++++++++++++++++++++++++++++++++++++++++++++++++++
\section{The fluid half-space}
%++++++++++++++++++++++++++++++++++++++++++++++++++

Adjoining the deformed solid half-space is a half-space $x_2 \le 0$
filled with an incompressible viscous Newtonian fluid, for which all
mechanical fields are denoted by a superscripted asterisk.  In the
static state the fluid is subject only to a hydrostatic stress
$\bm{\sigma}^*=-P^*\vec{I}$, and by continuity of traction across
the boundary $x_2=0$ we must have
\begin{equation}
-P^*=\sigma_2.
\end{equation}
The constitutive law for the fluid associated with the motion is
then written in terms of a superimposed Cauchy stress tensor,
denoted here by $\vec{s}^*$ and given by
\begin{equation}
\vec{s}^*=-p^*\vec{I}+2\mu^*\vec{D}^*,\quad \tr\vec{D}^*=0,
\end{equation}
where $\mu^*$ is the viscosity of the fluid,
\begin{equation}
\vec{D}^*=\frac{1}{2}[\grad\vec{v}^*+(\grad\vec{v}^*)^T],
\end{equation}
$\vec{v}^*$ is the fluid velocity, and $p^* = p^*(\vec{x},t)$.

We seek inhomogeneous waves with the same structure as in the solid,
that is
\begin{equation}
\{ \vec{v}^*, \vec{s}^*\}(x_1, x_2, x_3, t)
  = \{ \hat{\vec{v}}^*(x_2), -(k/\omega) \hat{\vec{s}}^*(x_2)\}
                \ee^{\ii k(c_\theta x_1 + s_\theta x_3)}
                  \ee^{- \ii \omega t}, \label{solnFluid}
\end{equation}
where $\hat{\vec{v}}^*$, $\hat{\vec{s}}^*$ are functions of $x_2$
only, such that
\begin{equation}
 \hat{\vec{v}}^*(-\infty) = \mathbf{0}, \quad
   \hat{\vec{s}}^*(-\infty) = \mathbf{0}.
\end{equation}
We find that the equations of motion, $\div\vec{s}^* = \rho^*
{\bf\dot{\vec{v}}}^*$ (where $\rho^*$ is the mass density of the
fluid), linearized in $\vec{v}^*$, can be cast as
\begin{equation}
  {\vec{\xi}^*}'(x_2) = \ii k \vec{N}^*\vec{\xi}^*(x_2),
\end{equation}
where
\begin{equation}
      \vec{\xi}^* =
    [\hat{v}^*_1, \hat{v}^*_2, \hat{v}^*_3,
     \hat{s}^*_{21}, \hat{s}^*_{22}, \hat{s}^*_{23}]^T,
\end{equation}
and the constant complex matrix $\vec{N}^*$ has the block structure
\begin{equation}
\vec{N}^*  =  \begin{bmatrix}
                    \vec{N}^*_1 \ & \vec{N}^*_2 \\[0.1cm]
   \vec{N}^*_3 + \hat{\rho}^* \vec{I}\ \
                      & {\vec{N}^*_1}
                             \end{bmatrix},
\end{equation}
$\vec{N}^*_1$, $\vec{N}^*_2$, $\vec{N}^*_3$ being real symmetric
matrices, and the notation $\hat{\rho}^*=\rho^* \omega^2 / k^2$ has
been adopted. If we write $\hat{\mu}^*=\mu^*\omega$, then,
respectively, $-\vec{N}^*_1$, $-\ii\hat{\mu}^*\vec{N}^*_2$, $
-\ii\vec{N}^*_3/\hat{\mu}^*$ are
\begin{equation}
 \begin{bmatrix}
       0 & c_\theta & 0 \\
       c_\theta & 0 & s_\theta \\
       0 & s_\theta & 0
       \end{bmatrix},
\quad
 \begin{bmatrix}
       1 & 0 & 0 \\
           0 & 0 & 0 \\
        0 & 0 & 1
      \end{bmatrix},
\quad
 \begin{bmatrix}
  4 c_\theta^2 + s_\theta^2 & 0
     & 3  c_\theta s_\theta \\
          0   &     0   &      0     \\
       3  c_\theta s_\theta & 0 &
     c_\theta^2 + 4s_\theta^2
                   \end{bmatrix}.\label{N1N2N3star}
\end{equation}

Again, when we seek a decaying partial-mode solution, this time in
the form
\begin{equation}
 \vec{\xi}^*(x_2)
 =  \ee^{ k s^* x_2} \vec{\zeta}^*,
 \quad \Re(k s^*) >0,
\end{equation}
where $\vec{\zeta}^*$ is a constant vector and $s^*$ an unknown
scalar, we end up with an eigenvalue problem, here
$\vec{N}^*\vec{\zeta}^* = -\ii s^* \vec{\zeta}^*$. The associated
\textit{propagation condition} is $\det (\vec{N}^* + \ii s^*
\vec{I}) =0$, which here simplifies to
\begin{equation}
(s^{\star 2} - 1)
  (s^{\star 2} - 1 + \ii \epsilon)^2
  = 0,
\end{equation}
with roots
\begin{equation}
 \pm 1, \quad
  \pm  \sqrt{1 - \ii \epsilon}\
\text{ (repeated)},\label{fluidRoots}
\end{equation}
where $\epsilon=\hat{\rho}^*/\hat{\mu}^* = \rho^* \omega/(\mu^*
k^2)$. The roots are independent of $\theta$, as expected, because
the fluid is isotropic. Corresponding to each of the four roots,
there are four eigenvectors and therefore potentially four
partial-modes. However, two of these must be discarded since their
amplitudes do not decay with distance from the interface $x_2=0$.
The two remaining modes form the basis for the general solution of
the equations of motion that is needed for matching with the
two-partial wave in the solid.

We now give the general boundary conditions at the deformed
solid/viscous fluid
interface.

%++++++++++++++++++++++++++++++++++++++++++++++++++
\section{The interface}
%++++++++++++++++++++++++++++++++++++++++++++++++++

In order to match the incremental tractions across the boundary it
is necessary to work in terms of the Cauchy stress since the nominal
stress is not defined inside the fluid.  Towards this end we first
calculate the incremental traction in the solid in terms of the
Cauchy stress.  Continuity of traction requires
\begin{equation}
\vec{S}^T\vec{N}{\rm d}A
  =\bm\sigma\vec{n}{\rm d}a
    =\bm\sigma^*\vec{n}{\rm d}a,
\end{equation}
where ${\rm d}A$ and ${\rm d}a$ are the area elements in
$\mathcal{B}_0$ and $\mathcal{B}$, respectively. Taking the
increment of this and updating the reference configuration to
$\mathcal{B}$ yields
\begin{equation}
\vec{s}^T\vec{n}{\rm d}a
  \equiv{\bf\tilde{\bm\sigma}}\vec{n}{\rm d}a
       + \bm{\sigma}\widetilde{\vec{n}{\rm d}a}
  = \vec{s}^*\vec{n}{\rm d}a
       + \bm{\sigma}^*\widetilde{\vec{n}{\rm d}a},
\end{equation}
where a superposed tilde indicates an increment. Note that, after
updating, $\vec{F}=\vec{I}$ and  $\vec{S}=\bm\sigma$ in the
configuration $\mathcal{B}$.

Now, according to Nanson's formula (applied to the boundary of the
solid), we have $\vec{n}{\rm d}a=\vec{F}^{-T}\vec{N}{\rm d}A$,
from which it follows, again after updating, that
\begin{equation}
\widetilde{\vec{n}{\rm d}a} = -(\grad \vec{u})^T\vec{n}{\rm d}a.
\end{equation}
Hence, the incremental traction continuity condition can be written
\begin{equation}
\vec{s}^T\vec{n} \equiv
 [{\bf\tilde{\bm\sigma}}-\bm\sigma(\grad\vec{u})^T]\vec{n}
   =[{\bf\tilde{\bm\sigma}}^*-\bm\sigma^*(\grad\vec{u})^T]\vec{n},
\end{equation}
and we recall that $\bm{\sigma}^*=-P^*\vec{I}$.

Since $\vec{n}$ is in the $x_2$ direction for the considered
half-space we may write the continuity condition in component form
as
\begin{equation}
s_{2i}=s^*_{2i}+P^*u_{2,i},\quad i=1,2,3,\quad
\mbox{on}\quad x_2=0.
\end{equation}
Additionally, the velocity must be
continuous, i.e.
\begin{equation}
v_i^*=v_i,\quad i=1,2,3,\quad \mbox{on}\quad
x_2=0.
\end{equation}

In terms of the functions $\hat{\vec{v}}(x_2),\hat{\vec{s}}(x_2)$
and their counterparts in the fluid, the boundary conditions become
\begin{equation}
\hat{v}_i^*(0)=\hat{v}_i(0),\quad i=1,2,3,\label{bc-v}
\end{equation}
and, noting that $v_i=-\ii\omega u_i$,
\begin{equation}
\hat{s}^*_{12}(0)+ c_\theta P^*\hat{v}_2(0)=\hat{s}_{21}(0),\quad
\hat{s}^*_{32}(0)+ s_\theta
P^*\hat{v}_2(0)=\hat{s}_{23}(0),\label{bc-s1}
\end{equation}
and
\begin{equation}
\hat{s}^*_{22}(0)-\frac{\ii}{k}
P^*\hat{v}_2'(0)=\hat{s}_{22}(0).\label{bc-s2}
\end{equation}

%++++++++++++++++++++++++++++++++++++++++++++++++++
\section{Principal waves: no restriction on $W$}
%++++++++++++++++++++++++++++++++++++++++++++++++++

%------------------------------------------------
\subsection{General solution in the solid}
%------------------------------------------------

Here we take $\theta=0$ and place no restriction on the form of $W$.
When $\theta = 0$, the in-plane mechanical fields in the solid
satisfy the equations of motion $\vec{\xi}' = \ii k \vec{N \xi}$,
where now $\vec{\xi}(x_2) = [\hat{v}_1, \hat{v}_2, \hat{s}_{21},
\hat{s}_{22}]^T$ and
\begin{equation} \label{Nprincipal}
\vec{N} = \begin{bmatrix}
  0\ &\ -1 + \overline{\sigma}_2\ &\ 1/\gamma_{21} & 0 \\
 -1\ &\ 0\ &\ 0 & 0 \\
  \hat{\rho} - \eta\
          &\ 0\ &\ 0 & -1 \\
  0\ &\   \hat{\rho}
           - \nu\
             &\    -1 + \overline{\sigma}_2\ & 0
          \end{bmatrix},
\end{equation}
with $ \overline{\sigma}_2 = \sigma_2/ \gamma_{21}$ and $\eta$ and
$\nu$ now reduced to
$$
\eta=2[\beta_{12}+\gamma_{21}(1-\overline{\sigma}_2)],\quad \nu =
\gamma_{12}+\gamma_{21}(1-\overline{\sigma}_2)^2.
$$

Here the propagation condition is a quadratic in $s^2$ \cite{Dowaikh90} and given by
\begin{equation}
 s^4 - 2\beta s^2 + \alpha^2 = 0,\label{quartic}
\end{equation}
where
\begin{equation}
2\beta = \dfrac{2 \beta_{12} - \hat{\rho}}
                  {\gamma_{21}},
\quad \alpha^2 =  \dfrac{\gamma_{12} - \hat{\rho}}
                     {\gamma_{21}}.
 \end{equation}
We recall that $\hat{\rho}=\rho\omega^2/k^2$. Formally, the roots
of the quartic \eqref{quartic} are
\begin{equation}
 \pm
\sqrt{\beta + \sqrt{\beta^2 - \alpha^2}}, \quad \pm \sqrt{\beta -
\sqrt{\beta^2 - \alpha^2}}.
  \label{rootsW}
\end{equation}

We pause the analysis to highlight a particular feature of the
present interface waves. Because the fluid is viscous, the wave
number $k$ is complex and so, therefore, are $\beta$ and $\alpha$.
It follows that, in contrast to the purely elastic case \cite{Dowaikh91}, 
it is not clear \emph{a priori} which two of
these four roots are such that the decay condition
Eq.~\eqref{solutionSolid}$_2$ is satisfied.  Let $s_1$ and $s_2$
be two such roots.  We note first that
\begin{equation}
s_1^2 + s_2^2 = 2\beta, \quad s_1^2 s_2^2 =
\alpha^2,\label{sumProduct}
\end{equation}
and then, depending on which value in Eq.~\eqref{rootsW} they
correspond to, one of the following four possibilities may arise:
 \begin{align}
   & s_1 s_2 = \alpha,
  &&\hspace*{-2.6cm} s_1 + s_2 = \sqrt{2(\beta + \alpha)},
\notag \\
   & s_1 s_2 = \alpha,
  &&\hspace*{-2.6cm} s_1 + s_2 = - \sqrt{2(\beta + \alpha)},
\notag \\
   & s_1 s_2 = -\alpha,
  &&\hspace*{-2.6cm} s_1 + s_2 =  \sqrt{2(\beta - \alpha)},
\notag \\
   & s_1 s_2 = -\alpha,
  &&\hspace*{-2.6cm} s_1 + s_2 = - \sqrt{2(\beta - \alpha)}.
\label{spW}
  \end{align}

Now, the eigenvectors $\vec{\zeta}^1$ and $\vec{\zeta}^2$ associated
with $s_1$ and $s_2$ are columns of the matrices adjoint to $\vec{N}
- \ii s_1 \vec{I}$ and $\vec{N} - \ii s_2 \vec{I}$, respectively.
Taking, for instance, the fourth column gives
\begin{equation}
\vec{\zeta}^i =
 [\ii s_i, -1, -\gamma_{21}(1 + s_i^2 - \overline{\sigma}_2),
    -\ii \gamma_{21}s_i(1 + 2 \beta - s_i^2 - \overline{\sigma}_2)]^T,
\quad i=1,2.
\end{equation}
Finally, the general solution in the solid is
\begin{equation} \label{gnlSolid}
 \vec{\xi}(x_2) = A_1 \ee^{- k s_1 x_2} \vec{\zeta}^1
                   + A_2 \ee^{- k s_2 x_2} \vec{\zeta}^2,
\end{equation}
where $A_1$ and $A_2$ are constants.

%------------------------------------------------
\subsection{General solution in the fluid}
%------------------------------------------------

In the fluid, a two-partial solution is required for adequate
matching at the interface. Taking $\theta = 0$ in Section 3, we
find that the in-plane equations of motion are ${\vec{\xi}^*}' =
\ii k \vec{N}^* \vec{\xi}^*$, where now $\vec{\xi}^*(x_2) =
  [\hat{v}^*_1, \hat{v}^*_2, \hat{s}^*_{21}, \hat{s}^*_{22}]^T$ and
\begin{equation}
\vec{N}^* = \begin{bmatrix}
  0 & -1\ & \ii/\hat{\mu}^* & 0 \\
 -1 & 0\ & 0 & 0 \\
 4\ii\hat{\mu}^* + \hat{\rho}^*  & 0\ & 0 & -1 \\
  0 &   \hat{\rho}^*\ & -1 & 0
          \end{bmatrix}.
\end{equation}

We assume that the wave propagates and is attenuated in the
direction $x_1 \ge 0$. From Eq.~\eqref{solnFluid} with $\theta =0$,
we see that these assumptions lead to
\begin{equation} \label{k}
 \Re(k) > 0, \quad \Im(k) > 0.
\end{equation}
On the other hand, the wave is also attenuated with distance from
the interface; it follows that we can discard the root $ - 1$ from
Eq.~\eqref{fluidRoots} and retain the root $ + 1$. The choice to
be made for the remaining two roots in Eq.~\eqref{fluidRoots} is
not so clear cut and for the time being we call $s^*$ the suitable
root; hence  $s^*$ is such that
\begin{equation}
s^{* 2} = 1 - \ii \epsilon, \quad \Re(k s^*)
>0. \label{s*}
\end{equation}
Finally, the general solution in the fluid is
\begin{equation} \label{gnlFluid}
 \vec{\xi^*}(x_2) =
    A^*_1 \ee^{k x_2}
     \begin{bmatrix}
       \ii \\
       1  \\
       -2\ii \hat{\mu}^* \\
       -\hat{\mu}^*(1 + s^{* 2})
     \end{bmatrix}
 + A^*_2 \ee^{ k s^* x_2}
     \begin{bmatrix}
      \ii s^*  \\
       1 \\
       -\ii\hat{\mu}^* (1 + s^{* 2})  \\
       -2 \hat{\mu}^* s^*
     \end{bmatrix},
\end{equation}
where $A^*_1$ and $A^*_2$ are constants.

%---------------------------------------------------
\subsection{Dispersion equation for the interface wave}
%---------------------------------------------------

When we specialize the general boundary conditions \eqref{bc-v},
\eqref{bc-s1} and \eqref{bc-s2} to the present context, we find a
linear homogeneous system of four equations for the four unknowns
$A_1$, $A_2$, $A^*_1$, $A_2^*$, for which the associated determinant
must be zero. After some manipulations, using
Eq.~\eqref{sumProduct}, we find that
\begin{equation} \label{disp_mat}
\begin{vmatrix}
 -s_1\ & -s_2\ & 1\ & s^* \\
 1\ & 1\ & 1\ & 1 \\
 \gamma_{21}(1 + s_1^2)\ &  \gamma_{21}(1 + s_2^2)\
   & -2\ii \hat{\mu}^*\ & -\ii\hat{\mu}^*(1 + s^{\star 2}) \\
 \gamma_{21}s_1(1 + s_2^2)\ &  \gamma_{21}s_2(1 + s_1^2)\
   & \ii \hat{\mu}^*(1 + s^{\star 2}) \ & 2\ii \hat{\mu}^* s^*
\end{vmatrix} =0.
\end{equation}
We see at once that the normal load $\sigma_2$ does not appear
explicitly in this equation. This feature highlights a major
difference between a wave at the interface of a loaded solid
half-space and a vacuum (Rayleigh wave) and a wave at the
interface of a loaded solid half-space and a viscous fluid, as
considered here; Chadwick and Jarvis \cite{Chadwick79a} also noted this
peculiarity for waves at the interface of two loaded solid
half-spaces (Stoneley wave). Of course, $\sigma_2$ still plays an
important role, in particular in the determination of the
pre-stretch ratios and of the amplitudes of the tractions in the
solid.

Once the determinant is expanded and the factors $(s_1 -
s_2)(1-s^*)$ are removed, we end up with the \textit{exact
dispersion equation},
\begin{multline}
\gamma_{21}^2 \left[
 \dfrac{\gamma_{12} - \hat{\rho}}{\gamma_{21}}
  + \dfrac{2 \beta_{12} + 2\gamma_{21} - \hat{\rho}}
           {\gamma_{21}}s_1 s_2 -1 \right]
 \\
  -\ii \gamma_{21}\hat{\mu}^* [2(1 - s_1 s_2)(1-s^*)
         + (s_1 + s_2)(1+s^*)(s^* + s_1 s_2)]
\\
    +
\hat{\mu}^{* 2} \left[
 \ii \epsilon
    +  \left(\ii \epsilon - 4 \right)s^* \right]
   = 0.\label{exact}
\end{multline}
This equation is fully explicit because the terms $s_1 + s_2$,
$s_1 s_2$, and $s^*$ are given by Eqs.~\eqref{spW} and
Eq.~\eqref{s*}. Of course, as noted earlier, there are several
possibilities for these terms, which generate in total eight
different dispersion equations. In each case the resulting root(s)
in $k$ must be checked for validity against the propagation and
decay conditions \eqref{solutionSolid}$_2$, \eqref{k}, and
\eqref{s*}$_2$, which we summarize here as
\begin{equation}
\label{conditions} \Re(k)
>0, \quad \Im(k) > 0, \quad \Re(k s^*) >0, \quad \Re(k s_1)
>0, \quad \Re(k s_2) >0.
\end{equation}
Notice that in the special case of solids whose strain-energy
function $W$ is such that $2 \beta_{12} = \gamma_{12} +
\gamma_{21}$, which includes the neo-Hookean solid
\eqref{neoHookean}, the biquadratic  \eqref{quartic} factorizes as
$(s^2 - 1)(s^2 - \alpha^2) = 0$. Hence $s_1 = 1$, $s_2 = \pm
\alpha$, and the exact secular equation \eqref{exact} simplifies
accordingly, leading this time to four different exact dispersion
equations.

%---------------------------------------------------
\subsection{Application: modelling of intravascular ultrasound}
%---------------------------------------------------

In recent years, intravascular ultrasound (IVUS) has proved to be
a most promising tool of investigation for measuring and assessing
abdominal aortic a\-neu\-rysms. Its accuracy is as good as that of
computed tomography (CT) scans and it has obvious non-radiative
advantages \cite{Lederle}, \cite{Wolf}, \cite{Garret}. 
We now apply the results of this section to an IVUS
context.

First we recall that medical ultrasound imaging devices operate in
the 1--10\,MHz range; accordingly we take $\omega = 10^7$\,Hz. We
argue, and will check \emph{a posteriori}, that at such high
frequency the wavelength is small compared with the radius and
thickness of an artery so that, as far as the propagation of
localized waves is concerned, an aortic aneurysm can be modelled
as a half-space. Here we take the axis of the artery along the
$x_1$ direction and consider that the half-space $x_2 \le 0$ is
filled with blood.

For the solid, we use a strain-energy function devised by Raghavan
and Vorp \cite{Raghavan} to fit experimental data collected on uniaxial
tension tests of aortic aneurysms, namely
\begin{equation} \label{Vorp}
W = \mathcal{C}_1 (\lambda_1^2 + \lambda_2^2 +\lambda_3^2 -3)
    + \mathcal{C}_2 (\lambda_1^2 + \lambda_2^2 +\lambda_3^2 -3)^2,
\end{equation}
where, typically, $\mathcal{C}_1 = 0.175$ MPa, $\mathcal{C}_2 =
1.9$ MPa. We assume that the aneurysm corresponds to a region
where the tissue undergoes a strain of 20\%, and that the
end-systolic blood pressure is high, 150\,mmHg ($=20$\,kPa) say,
so we take $\sigma_2=20$\,kPa.
We consider that the tissue is
free to expand or contract in the $x_3$ direction with
$\sigma_3=0$. From \eqref{initialStress} we find the remaining
stresses and strains. Summarizing, we have
\begin{equation}
 \lambda_1 = 1.2, \quad \lambda_3 \simeq 0.908,\quad
 \sigma_1 \simeq 71.45\,\mbox{kPa},\quad
  \sigma_2 = 20\,\mbox{kPa},\quad
 \sigma_3 = 0,
\end{equation}
with $\lambda_2$ calculated from the incompressibility condition.
For the mass density, we take $\rho = 1000$ kg/m$^3$. For the
blood, we use typical values of viscosity \cite{Spring}
and mass density \cite{Kenner}: $\mu^* = 3.5 \times
10^{-3}$\,Ns/m$^2$, $\rho^* = 1050$\,kg/m$^3$.

Here, the frequency $\omega$ is fixed as a real quantity \textit{a
priori}.  We then replace $k$ everywhere by $k = \omega S$, where
$S= S^+ + \ii S^-$ is the (complex) \textit{slowness}, and the
only unknown in the dispersion equation \eqref{exact}. The
propagation and decay conditions \eqref{conditions} are satisfied
and may be written as
\begin{equation}
\label{conditions_S}
\Re(S) >0, \quad
\Im(S) > 0, \quad
\Re(S s^*) >0, \quad
\Re(S s_1)>0, \quad
\Re(S s_2) >0.
\end{equation}

We find that the only qualifying root is
 \begin{equation}
 S =  2.721 \times 10^{-2} + 4.424 \times 10^{-4} \ii,
\end{equation}
from which we deduce the phase speed $v =
1/S^+$, the damping factor $\gamma = S^-$, and
 the wavelength $\lambda_0 = 2 \pi / (\omega S^+)$:
\begin{equation}
v = 36.75\,\mbox{m/s}, \quad \gamma =  4.424 \times 10^{-4} \,\mbox{m}^{-1},
\quad \lambda_0 = 23.1\,\mu\text{m}.
\end{equation}

Then we plot the depth profiles of the wave. Its amplitude is the
real part of $\vec{v}$ in the solid and of $\vec{v}^*$ in the
fluid. Explicitly,
\begin{multline}
\Re \left(\vec{v}(x_1, x_2, t)\right) =
 \ee^{- \gamma x_1} \left[
  \Re\left(\hat{\vec{v}}(x_2)\right) \cos \omega (x_1 / v - t) \right].\\
   \left. - \Im\left(\hat{\vec{v}}(x_2)\right) \sin \omega (x_1 / v - t)
  \right],
\end{multline}
in the solid, and similarly for the fluid, for which $\vec{v}$ and
$\hat{\vec{v}}$ are replaced by their asterisk counterparts.
Clearly, the particle velocity is elliptically polarized, and the
lengths of the ellipse semi-axes decay with distance away from the
interface and also with increasing $x_1$. Figure 1 shows the
variations of the normal (continuous curve) and tangential (dotted
curve) velocity components in the fluid ($x_2 \le 0$) and in the
solid ($x_2 \ge 0$), normalized with respect to $\hat{v}_2(0)$, as
functions of $x_2 / \lambda_0$. Note that the components are in
phase quadrature. Note also, as is clear from the zooms shown in
Figure 1, that the components are continuous across the interface,
as expected, but their first derivatives are discontinuous. The
wave is elliptically polarized near the interface; the major axis
is normal to the interface and more than 12 times the length of
the minor axis; the ellipse is described in a retrograde manner.
This shape is carried through the depth of the solid whereas in
the fluid, the wave becomes rapidly circularly polarized at a
depth of about 0.06 wavelengths, and remains nearly so through the
rest of the half-space. The localization is greater in the fluid
than in the solid: the amplitude has almost vanished after one
wavelength into the former and after five wavelengths in the
latter. An aneurysm is typically 1\,mm thick, which, with the
numerical values used here, is more than 50 wavelengths; thus, the
assumption of a semi-infinite solid is justified, as is the
assumption of a flat interface (aneurysms are typically of
diameter 5\,cm).

\begin{figure}[!t]
\centering
  \epsfig{figure=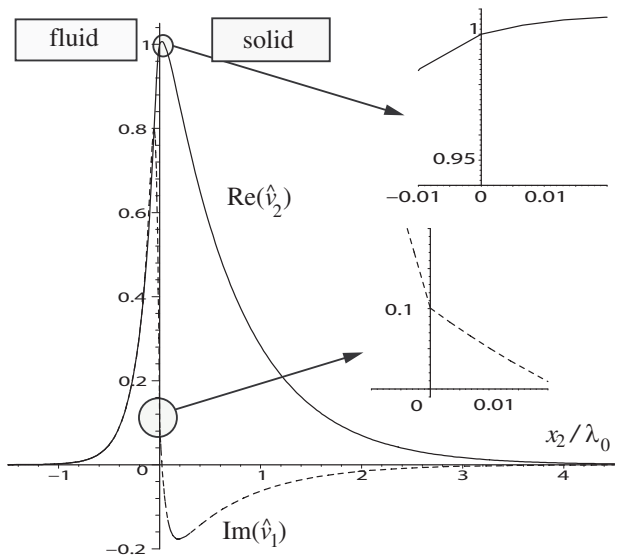,
width=.6\textwidth}
 \caption{Depth profiles as functions of $x_2/\lambda_0$ of the
velocity components of the acoustic wave: $x_2$ (normal) component
-- continuous curve; $x_1$ (tangential) component -- dotted curve.
The zooms show continuity of these components and discontinuity of
their derivatives.}
\end{figure}

Finally, we note that if the fluid is absent then the
corresponding surface wave would travel with speed 40.734\,m/s;
hence the viscous fluid not only dampens the wave but also slows
it down noticeably. Of course, blood flows in an artery, and
creates shear deformation and stress in the solid. The influence
of wall shear stress on the waves will be treated elsewhere.

%---------------------------------------------------
\subsection{Example: compressive stresses}
%---------------------------------------------------

Here we consider the behaviour of the elastic half-space (in
contact with the viscous fluid) when it is compressed in the $x_1$
direction ($\lambda_1 <1$). In their pioneering works, Green and
Zerna \cite{Green} and Biot \cite{Biot63}, \cite{Biot65} 
showed that when a highly elastic half-space with a
free surface (no fluid contact) is compressed, a surface
instability may develop. Focusing on neo-Hookean solids
\eqref{neoHookean} they showed that the critical stretches are
$\lambda_\text{cr} = 0.666$ for tangential equi-biaxial
compression ($\lambda_1 = \lambda$, $\lambda_2 = \lambda^{-2}$,
$\lambda_3 = \lambda$), $\lambda_\text{cr} = 0.544$ for plane
strain compression ($\lambda_1 = \lambda$, $\lambda_2 =
\lambda^{-1}$, $\lambda_3 = 1$), and $\lambda_\text{cr} = 0.444$
for normal equi-biaxial compression ($\lambda_1 = \lambda$,
$\lambda_2 = \lambda^{-1/2}$, $\lambda_3 = \lambda^{-1/2}$).
Theirs was a static stability analysis, later also included in a
wider dynamical context by Flavin \cite{Flavin}, Willson \cite{Willson}, Chadwick
and Jarvis \cite{Chadwick79b}, Dowaikh and Ogden \cite{Dowaikh90}, and others. We now
consider this problem.

A localized small-amplitude wave propagates over the free surface
of a deformed Mooney-Rivlin or neo-Hookean half-space with
normalized squared speed $\rho v^2 / \gamma_{12} = 1 - \sigma_0^2
\lambda_2^2/\lambda_1^2$, where $\sigma_0$ is the real root of
$\sigma^3 + \sigma^2 + 3 \sigma -1 = 0$ ($\sigma_0 = 0.2956$).
Clearly, in the examples of plane strain and equi-biaxial strain
above, the squared wave speed increases when $\lambda$ increases
and decreases when $\lambda$ decreases. At $\lambda = 1$ there is
no pre-strain and $\rho v^2 / C = 0.9126$, the value found by Lord
Rayleigh \cite{Rayleigh} for isotropic incompressible linearly elastic
solids. As $\lambda \rightarrow \infty$, the squared wave speed
tends to the squared wave speed $\gamma_{12}/ \rho$ of a
transverse bulk wave. As $\lambda$ decreases, there is a critical
stretch $\lambda_\text{cr}$ at which the squared speed is zero and
below which $v^2 < 0$. Since the wave time dependence is of the
form $\ee^{\ii k (x_1 - v t)}$, with $k>0$, it follows that the
amplitude grows without bound in time when $\lambda <
\lambda_\text{cr}$, and that the surface becomes unstable (at
least, in the linearized theory). It is therefore appropriate to
ask what happens when the compressed half-space is in contact with
a viscous fluid.

Bagno and co-workers addressed this question in a series of
articles (see the review by Bagno and Guz \cite{Bagno97} and references
therein). For a neo-Hookean solid they found that the wave speed
drops to zero when $\lambda = 0.544$ in plane strain compression
and when $\lambda = 0.444$ in normal equi-biaxial compression,
that is at the \emph{same} critical stretches as for surface
(solid/vacuum) instability, irrespective of the viscous fluid
characteristics. On the other hand, Bagno and Guz \cite{Bagno87} find that
the wave speed falls to zero at a critical stretch that
\emph{does} depend upon the viscosity of the fluid. To address
this disparity, we now compute the speed of the interfacial wave
when the half-space is in compression.

In order to minimize the number of parameters, we use the
neo-Hookean solid, with $W$ given by \eqref{neoHookean}. We take a
normal equi-biaxial pre-strain ($\lambda_1 = \lambda$, $\lambda_2
= \lambda^{-1/2}$, $\lambda_3 = \lambda^{-1/2}$) and, following
Bagno and Guz, we take the frequency $\omega$ to be real. A
dimensional analysis of the resulting dispersion equation shows
that \eqref{exact} now depends on just three non-dimensional
parameters: a measure of the pre-strain, $\lambda$; a measure of
the dynamic viscosity of the fluid compared with the shear modulus
of the solid, $\mu^* \omega / C$; and the ratio of the densities,
$\rho^* / \rho$. Once these quantities are specified, the
dispersion equation may be solved for the non-dimensional complex
unknown $x$ defined by
\begin{equation}
x:= \sqrt{\dfrac{\rho}{\gamma_{12}}} \dfrac{\omega}{k} =
\sqrt{\dfrac{\rho}{C \lambda^2}} S^{-1},\label{x-k}
\end{equation}
where $S$ is the (complex) scalar slowness.
The dispersion equation can now be solved numerically for $x$,
and the interfacial wave speed,
normalized with respect to the transverse bulk shear wave
speed in the deformed solid, is $c = \Re(1/x)$.

For Figure 2(a), we fix $\rho^* / \rho$ at 1.0 and we take in turn
$\mu^* \omega / C =$ 0.2, 0.04, 0.02. The first choice ($\rho^* /
\rho =$ 1.0, $\mu^* \omega / C =$ 0.2) is roughly that obtained
for the blood/artery interface of the previous section with
$\mathcal{C}_1 = C$ and $\mathcal{C}_2 = 0$. We see clearly that
as $\mu^* \omega / C$ decreases the wave speed decreases towards
zero as $\lambda$ tends to 0.444, the critical compression stretch
for the solid/vacuum interface (the thick curve gives the
solid/vacuum interface wave speed).  We find that in the extension
to moderate compression range, the solid/fluid interface wave
speed is significantly lower than the solid/vacuum wave speed. In
the strongly compressive range (as $\lambda \rightarrow 0.444$),
the speed plot dips towards zero, and dips further as $\mu^*
\omega / C$ decreases, \textit{without ever reaching that value};
the plot gets close to the plot for the solid/vacuum interface
wave speed as $\lambda$ decreases, but then crosses it, and the
fluid/solid interface wave speed increases again. Note that we
checked the validity of the solution at all compressive stretches
using \eqref{conditions_S}.

For Figure 2(b), we fixed $\mu^* \omega / C$ at 1.0 and took
$\rho^*/\rho = 0.1, 0.05, 0.01$. In the extension to moderate
compression range, the plot for the solid/fluid interface wave
speed gets closer to that for the solid/vacuum interface wave
speed as $\rho^* / \rho$ decreases. In the strongly compressive
range, similar comments to those made for Figure 2(a) apply.

\begin{figure}[!t]
  \epsfig{figure=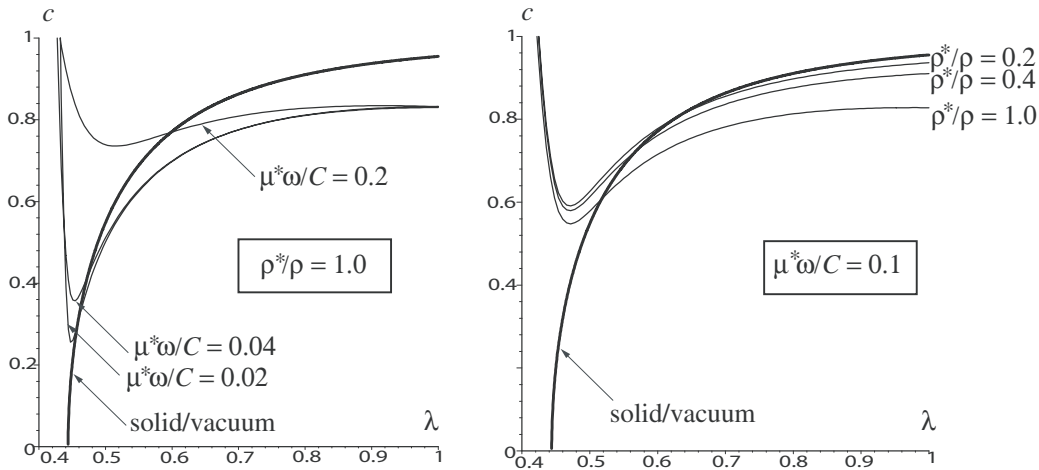, width=\textwidth}
 \caption{Normalized wave speed for a neo-Hookean
 solid under normal equi-biaxial pre-strain and a viscous fluid: (a) $\rho^* / \rho=$
 1.0 and $\mu^* \omega / C =$ 0.2, 0.04, 0.02;
 (b)  $\rho^* / \rho=$
 1.0, 0.4, 0.2 and $\mu^* \omega / C =$ 0.1. The thick curves are
 for an unloaded solid half-space.}
\end{figure}

The conclusion is that as both $\mu^* \omega / C$ and
$\rho^*/\rho$ tend to zero, the speed tends to zero when the
stretch tends to the critical compression stretch of the
solid/vacuum surface instability. This is to be expected because
this double limit corresponds to the vanishing of the fluid.
However, as emphasized earlier, the speed never reaches the zero
limit and the bifurcation criterion is therefore never met. For
instance, the typical values used by Bagno and co-workers \cite{Bagno97}
put $\mu^* \omega / C$ at about 0.0002 and
$\rho^*/\rho$ as low as 0.1, giving $c$ smaller than $10^{-4}$,
but not zero. In effect a localized damped wave exists for the
whole compressive range, with speed values starting below the
solid/vacuum interface wave speed in the moderate compressive
range, reaching a minimum (above the solid/vacuum interface wave
speed) as $\lambda$ approaches 0.444, and then rising rapidly to
infinity as $\lambda$ decreases below 0.444. This result suggests
that when a neo-Hookean half-space is in contact with a viscous
fluid it becomes completely stable. This, however, is an incorrect
deduction, because it is based on special motions, for which the
frequency $\omega$ is assumed real. Other motions might be
unstable, as is illustrated below.

We now take the wave number $k$ to be real ($k>0$) and let
the speed $v$ be complex,
\be
k>0, \qquad
v = v^+ + \ii v^-,
\en
so that the motion is now proportional to
$\ee^{k v^- t} \ee^{\ii k (x_1 - v^+ t)}$.
Clearly in this case, the conditions for a stable, localized
wave, propagating in the $x_1>0$ direction, are
\begin{equation}
\label{conditions_v}
\Re(v) \ge 0, \quad
\Im(v) \le 0, \quad
\Re(s^*) >0, \quad
\Re(s_1)>0, \quad
\Re(s_2) >0.
\end{equation}
(note that $ \Re(s_1)>0$ is automatically satisfied in a
neo-Hookean solid, because $s_1 = 1$).

Then, an analysis of the dispersion equation \eqref{exact}, in the
case of a neo-Hookean solid with normal equi-biaxial pre-strain,
reveals three non-dimensional quantities: $\lambda$, $\mu^* k /
\sqrt{\rho C}$, and $\rho^* / \rho$. Once they are specified, we
solve the dispersion equation for the non-dimensional quantity $x$
defined as \be x := \sqrt{\dfrac{\rho}{\gamma_{12}}}
\dfrac{\omega}{k}
 = \sqrt{\dfrac{\rho}{C\lambda^2}} v.
\en The interfacial wave speed, normalized with respect to the
transverse bulk shear wave speed in the deformed solid, is $c =
\Re(x)$.

For Figure 3, we take $\rho^* / \rho = 1.0$, and $\mu^* k /
\sqrt{\rho C} =  0.2,  0.002$ in turn, and we plot $v$ against
$\lambda$ in the compressive range. We find that at a compressive
stretch close to the critical compressive stretch of surface
stability for the solid/vacuum interface, the normalized speed $c$
drops to zero.  From the figure it is not clear that the values of
$\lambda$ at this point are different for 0.2 and 0.002, but the
zoom in Figure 3 shows, however, that the value of the compressive
stretch at which $c=0$ depends on the material parameters. For
comparison, the curve corresponding to $\mu^* k / \sqrt{\rho C}
=1$ is also shown.  We emphasize that the situation at $c=0$ does
not correspond to a static solution of the equations of motion
(which would be impossible in the fluid), but to a non-propagating
damped motion, proportional to $\ee^{k v^- t} \ee^{\ii k x_1 }$.
Moreover, this situation does not correspond to an instability
because at that point, and a bit beyond, the requirements
\eqref{conditions_v} still hold. For instance, in the case $\rho^*
/ \rho = 1.0$, $\mu^* k / \sqrt{\rho C} = $ 0.2, the speed drops
to zero at $\lambda \simeq 0.44539$ but there still exist
non-propagating, localized motions for $\lambda > 0.42212$. Beyond
that stretch value, however, all solutions of the considered type
grow unboundedly with time and/or with space, indicating
instability. From this example we see that the viscous load
stabilizes slightly the solid half-space, because it can now be
compressed by an extra 2\%, from 0.444 to 0.422.

\begin{figure}[!t]
  \epsfig{figure=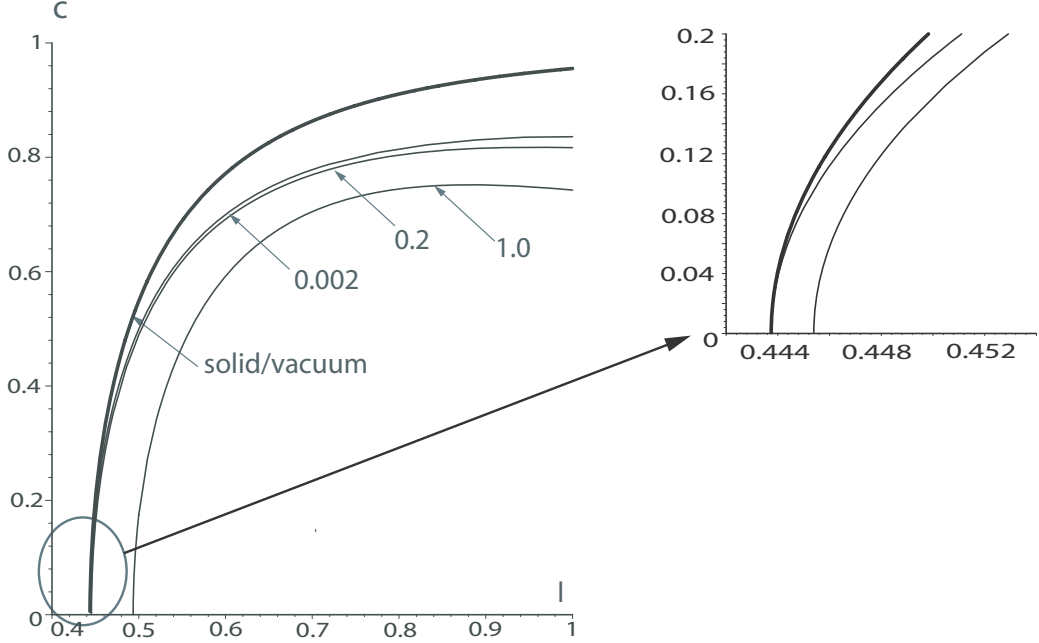, width=\textwidth}
 \caption{Normalized wave speed for a neo-Hookean
 solid under normal equi-biaxial pre-strain and a viscous fluid
 with $\rho^* / \rho = 1.0$ and
$\mu^* k / \sqrt{\rho C} =  0.2,  0.002, 1$ (indicated by the
arrows) for the compressive range $0.4 < \lambda < 1$.  The zoom
shows the details for $\mu^* k / \sqrt{\rho C} =  0.2,  0.002$ for
$0.442 < \lambda < 0.445$ range.  The thick curve corresponds to
the solid half-space with no fluid loading.}
\end{figure}

%++++++++++++++++++++++++++++++++++++++++++++++++++++
\section{Non-principal wave: neo-Hookean solid}
%++++++++++++++++++++++++++++++++++++++++++++++++++++

In this section we consider that the solid half-space is a
deformed neo-Hookean material, with strain-energy function given
by Eq.~\eqref{neoHookean}. For this material, Flavin \cite{Flavin}
noticed that Rayleigh waves are plane-polarized for any direction
of propagation in a principal plane; see \cite{Braun}, \cite{Dowaikh}. 
Chadwick and Jarvis \cite{Chadwick79a} showed the same
result for Stoneley waves. Indeed, we now show that the saggital
motion is \textit{always decoupled} from the `anti-saggital'
motion, for any $\theta$ and any type of (bulk or interfacial)
inhomogeneous wave.

First, we note that for neo-Hookean bodies, the parameters
$\gamma_{ij}$ and $\beta_{ij}$ have the simple expressions
\begin{equation}
 \gamma_{ij}  = C \lambda_i^2, \quad
  \beta_{ij}  = C (\lambda_i^2 + \lambda_j^2)/2.
\end{equation}
It follows that the matrices $\vec{N}_1$, $\vec{N}_2$, $\vec{N}_3$,
given by Eq.~\eqref{N1N2N3}, are greatly simplified. Next, we recall
that the direction of propagation and the normal to the interface
define the \textit{saggital plane}, with unit normal $[-s_\theta, 0,
c_\theta]^T$. Now consider the new unknown functions $w_i$, $t_{2i}$
($i=1, 2, 3$), defined by
\begin{equation}
 w_i = \Omega_{ij}\hat{v}_j, \quad
 t_{2i} = \Omega_{ij} \hat{s}_{2j},
 \end{equation}
 where
\begin{equation}
 \Omega_{ij} = \begin{bmatrix}
                    c_\theta & 0 & s_\theta \\
                    0 & 1 & 0 \\
                   -s_\theta & 0 & c_\theta
                \end{bmatrix}.
\end{equation}
Some algebraic manipulations reveal that the equations of motion
\eqref{1stOrder}, written for $w_i$, $t_{2i}$, decouple the
`anti-saggital' motion $[w_3, t_{23}]$ from its saggital
counterpart. For the latter we find
\begin{equation}
[w_1', w_2', t_{21}', t_{22}']^T
 =  \ii k \vec{N}[w_1, w_2, t_{21}, t_{22}]^T,
\end{equation}
with $\vec{N}$ in the form \eqref{Nprincipal}, where now
\begin{align}
& \eta =
   C(c_\theta^2 \lambda_1^2 + s_\theta^2 \lambda_3^2  + 3 \lambda_2^2)
     - 2\sigma_2,
&& \gamma_{21} = C \lambda_2^2,
\notag \\[0.1cm]
& \nu =
   C(c_\theta^2 \lambda_1^2 + s_\theta^2 \lambda_3^2)
   + C \lambda_2^2(1 - \overline{\sigma}_2)^2,
&& \overline{\sigma}_2 = \sigma_2 / (C \lambda_2^2).
\end{align}

A search for partial-mode solutions in the form $[w_1, w_2,
t_{21}, t_{22}]^T = \vec{\zeta} \ee^{- k s x_2}$, where
$\vec{\zeta}$ is a 4-component constant vector, leads to an
eigenvalue problem. The associated characteristic equation is the
propagation condition
\begin{equation}
 (s^2 - 1)[C \lambda_2^2 s^2 - C(c_\theta^2 \lambda_1^2 + s_\theta^2
\lambda_3^2 ) + \hat{\rho}] = 0,
\end{equation}
with roots $s = -1$, which is discarded because it does not lead
to a decaying wave,
\begin{equation} \la{s1s2} s_1 = 1, \quad s_2
= \pm \sqrt{[C(c_\theta^2 \lambda_1^2 + s_\theta^2 \lambda_3^2 )
                                     - \hat{\rho}]/(C \lambda_2^2)}.
\end{equation}
The analysis leading to the derivation of the dispersion equation
is by and large the same as that conducted for the principal wave
in Section 5. The end result is that the dispersion equation is
again \eqref{disp_mat}, where now $s_1$ and $s_2$ are given by the
expressions above, and $\gamma_{21}$ is replaced by
$C\lambda_2^2$, but the other quantities remain unchanged.

As an illustration, we take the same numerical values as in
Section 5.4, but with $\mathcal{C}_2 = 0$ in \rr{Vorp}. Hence the
solid is neo-Hookean with $C = 0.175$\,MPa and $\rho =
1000$\,kg/m$^3$. It is under the pre-stress $\sigma_2 = 20$\,kPa,
$\sigma_3 = 0$. The fluid has Newtonian viscosity $\mu^* = 3.5
\times 10^{-3}$\,Ns/m$^2$, and mass density $\rho^* =
1050$\,kg/m$^3$. The frequency is $\omega=10^7$\,Hz.
In turn, we take the
solid to be compressed by $20 \%$ in the $x_1$ direction (so that
$\lambda_1 = 0.8$ and then $\lambda_2 \simeq 1.131$, $\lambda_3
\simeq 1.105$), to be unstretched in the $x_1$ direction (so that
$\lambda_1 = 1$, and $\lambda_2 \simeq 1.014$, $\lambda_3 \simeq
0.986$), and to be under an extension of $20 \%$ in the $x_1$
direction (so that $\lambda_1 = 1.2$, and $\lambda_2 \simeq
0.929$, $\lambda_3 \simeq 0.897$).

Figure 4 shows the dependence of the interfacial wave speed $v^+ =
\Re{(\omega / k)}$ on the angle of propagation $\theta$. When the
solid is almost unstrained, the speed hardly varies with the
angle; when it is strained to $\pm 20 \%$, the induced anisotropy
causes speed changes of more than $\pm 25 \%$ in some directions.
The figure also shows clearly that the wave travels at its fastest
in the direction of greatest stretch, and at its slowest in the
direction of greatest compression, indicating that it could be
used for the acoustic determination of these directions.
\begin{figure}
 \centering
  \epsfig{figure=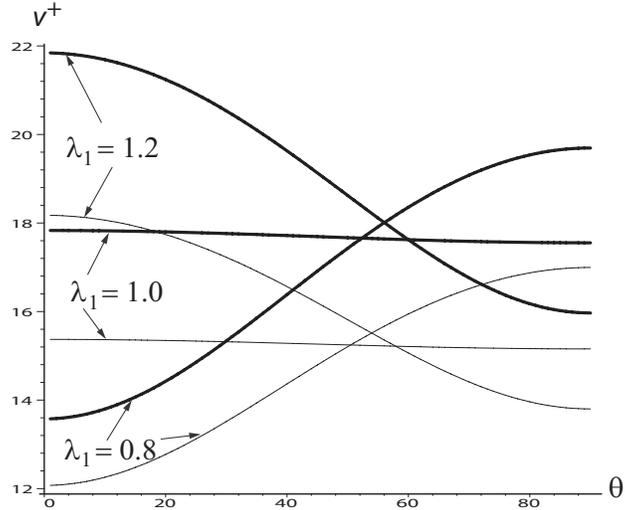, height=.5\textwidth, width=.6\textwidth}
\caption{Influence of propagation angle $\theta$ on the speed
$v^+$ of a wave at the interface between a viscous fluid and a
deformed neo-Hookean solid (thin curves).
The thick curves represent the wave speed in the absence of fluid loading.}
\end{figure}

%+++++++++++++++++++++++++++++++++++++++++++++++++++++
% bibliography
%++++++++++++++++++++++++++++++++++++++++++++++++++++++

\end{document}